\documentclass[%
superscriptaddress,
 amsmath,amssymb,
 aps,
prc,reprint,nofootinbib, showpacs,showkeys
]{revtex4-1}
\pdfoutput=1
\usepackage{graphicx}
\usepackage{dcolumn}
\usepackage{bm}
\usepackage{slashed}
\usepackage{graphicx}
\usepackage[colorlinks=true,linktocpage=true,linkcolor=blue,citecolor=blue]{hyperref}
\usepackage[utf8]{inputenc}
\usepackage[T1]{fontenc}    
\usepackage{float}
\usepackage{nicefrac}
\usepackage[normalem]{ulem}
\usepackage{amsmath}
\usepackage{amstext}
\usepackage{subfigure} 
\usepackage{bbold} 
\usepackage[makeroom]{cancel}
\usepackage{stmaryrd} 
\usepackage{makecell}
\newcolumntype{R}{>{$}r<{$}}
\newcolumntype{C}{>{$}c<{$}}
\usepackage{colortbl}
\definecolor{jocol}{rgb}{0.1,0.2,0.8}

\definecolor{tgcol}{rgb}{0.7,0.1,0.2}

\newcommand{\beq}{\begin{eqnarray}}
\newcommand{\eeq}{\end{eqnarray}}

\def\LR{\left(} 
\def\RR{\right)}
\def\LS{\left[} 
\def\RS{\right]}
\def\LC{\left\{} 
\def\RC{\right\}}

\def\nn{\nonumber}

\usepackage{lineno}
\begin{document}
\title{Spheroidal expansion and freeze-out geometry of heavy-ion collisions \\in the few-GeV energy regime}  

	\author{Szymon Harabasz}
	\email{s.harabasz@gsi.de}
	\affiliation{Technische Universit\"at Darmstadt, 64289 Darmstadt, Germany}
	
    \author{J\k{e}drzej Ko\l{}a\'{s}}
	\email{jedrzej.kolas.dokt@pw.edu.pl}
	\affiliation{Warsaw University of Technology, 00-662 Warsaw, Poland}

    \author{Radoslaw Ryblewski} 
    \email{radoslaw.ryblewski@ifj.edu.pl} 
    \affiliation{Institute of Nuclear Physics Polish Academy of Sciences, PL-31342 Krakow, Poland}

	\author{Wojciech Florkowski}
	\affiliation{Jagiellonian University,  PL-30-348 Krak\'ow, Poland}
	 
	\author{Tetyana Galatyuk}
	\affiliation{GSI Helmholtzzentrum f\"ur Schwerionenforschung GmbH, 64291 Darmstadt, Germany}
	\affiliation{Technische Universit\"at Darmstadt, 64289 Darmstadt, Germany}
	 
	\author{Ma\l{}gorzata Gumberidze}
	\affiliation{GSI Helmholtzzentrum f\"ur Schwerionenforschung GmbH, 64291 Darmstadt, Germany}

	\author{Piotr Salabura}
	\affiliation{Jagiellonian University,  PL-30-348 Krak\'ow, Poland}
	 
	\author{Joachim Stroth}
	\affiliation{GSI Helmholtzzentrum f\"ur Schwerionenforschung GmbH, 64291 Darmstadt, Germany}
    \affiliation{Institut f\"ur Kernphysik, Goethe-Universit\"at, 60438 Frankfurt, Germany}

    \author{Hanna Paulina Zbroszczyk}
    \affiliation{Warsaw University of Technology, 00-662 Warsaw, Poland}

\date{\today} 
\bigskip
\begin{abstract}
A spheroidal model of the expansion of hadronic matter produced in heavy-ion collisions in the few-GeV energy regime is proposed. It constitutes an extension of the spherically symmetric Siemens-Rasmussen blast-wave model used in our previous works. The spheroidal form of the expansion, combined with a single-freeze-out scenario, allows for a significantly improved description of both the transverse-mass and the rapidity distributions of the produced particles. With the model parameters determined by the hadronic abundances and spectra,  we make further predictions of the pion HBT correlation radii that turn out to be in a qualitative agreement with the measured ones. The overall successful description of the data supports the concept of spheroidal symmetry of the produced hadronic systems in this energy range.
\end{abstract}
     
\date{\today}  
\keywords{heavy-ion collisions, statistical hadronization, thermalization, hadron spectra, thermal model}
\maketitle 
%
\section{Introduction}
Thermal models of hadron production, based on the idea of statistical hadronization, have been very successful in describing hadron yields and phase-space distributions in various collision processes, especially in ultrarelativistic heavy-ion collisions (UrHIC) in a wide range of beam energies and for different projectile-target systems~(see, e.g., Refs.~\cite{Cleymans:1992zc,Becattini:1997ii,Florkowski:2001fp,Petran:2013lja,Becattini:2014hla,Vovchenko:2015idt,Andronic:2017pug,Mazeliauskas:2018irt,Castorina:2019vex}). The reasons for studying thermal aspects of hadron production in heavy-ion collisions are manifold. The hadron abundances can be explained over several orders of magnitude of multiplicity by fixing a small number of thermodynamic parameters. Moreover, the assumption of local thermalization of the expanding dense and hot matter formed in the collision (called a \emph{fireball}) allows for the application of  hydrodynamic concepts  \cite{Florkowski:2010zz,Florkowski:2017olj,Romatschke:2017ejr,Brachmann:1997bq} to describe its  evolution and emissivity of electromagnetic radiation~\cite{Adamczewski-Musch:2019byl}. Such an approach has been very successful in describing UrHIC and helped to identify landmarks in the phase diagram of Quantum Chromodynamics (QCD) in the region of vanishing net-baryon density, which is also accessible by lattice-QCD calculations~\cite{Borsanyi:2010bp,Bazavov:2018mes}. 

Heavy-ion collisions (HIC) at lower beam energies provide access to the strongly interacting matter at high net-baryon densities where a rich structure in the QCD phase diagram is predicted \cite{Isserstedt:2019pgx,Gao:2020fbl}, but lattice QCD is not applicable.
The question of whether the fireball formed in the few-GeV beam energy range (where in head-on collisions essentially all nucleons are stopped in the center-of-mass) is thermalized remains still a matter of debate~\cite{Lang:1990zc,Endres:2015fna,Galatyuk:2015pkq,Staudenmaier:2017vtq}. The study of hadron spectra is crucial to answering this question. However, in a thermal analysis it must first be demonstrated that the experimental hadron yields can be well described with a few thermodynamic parameters such as temperature, $T$, and the baryon chemical potential, $\mu_{B}$. 
Only in the second step the transverse-mass spectra, which are typically falling off exponentially, have to be reproduced. 
One should recall that collective radial expansion (specified by the radial velocity~$v$) and resonance decays also affect the momentum distribution of hadrons~\cite{Florkowski:2001fp,Schnedermann:1993ws,Broniowski:2001we}. 

The two physical aspects mentioned above are unified in a \emph{single-freeze-out model}~\cite{Broniowski:2001we,Broniowski:2002wp}, which identifies the chemical and kinetic freeze-outs by neglecting hadron rescattering processes (after the chemical freeze-out). This model assumes a sudden freeze-out governed by local thermodynamic conditions. This concept is implemented in the \textsc{THERMINATOR} Monte-Carlo generator~\cite{Kisiel:2005hn,Chojnacki:2011hb}, which allows for studies of hadron production taking place on arbitrary freeze-out hypersurfaces defined in the four-dimensional space-time. The most popular parametrization of such a freeze-out hypersurface~\cite{Schnedermann:1993ws}, dubbed the \emph{blast-wave model}, assumes the symmetry of boost invariance (along the beam axis) which was observed in UrHIC. In fact, it was introduced as a modification of the original blast-wave model formulated by Siemens and Rasmussen (SR)~\cite{Siemens:1978pb}, which instead of the boost invariance, employed a spherical symmetry of the freeze-out geometry~\cite{Florkowski:2010zz}. 

In our previous work \cite{Harabasz:2020sei}, using the SR model we introduced a novel approach toward a consistent simultaneous description of hadron yields and transverse-mass spectra. This framework offered an alternative interpretation of experimental results in the analyzed energy domain, as it was based on the thermal equilibrium concept, as compared to commonly used non-equilibrium transport model approaches~\cite{Bass:1998ca,Buss:2011mx,Petersen:2018jag,Hartnack:1997ez,Cassing:1999es}.  We assumed the spherical symmetry of a fireball to be a natural approximation at low energies, where the colliding nuclei are not transparent to each other (the energy dependence of this effect is shown in Ref.~\cite{Bearden:2003hx} and recently also systematical investigated in terms od stopping in: Ref.~\cite{Braun-Munzinger:2020jbk}). The agreement between the transverse-mass spectra of particles predicted by the model and measured by the HADES collaboration in Au-Au collisions at $\sqrt{s_{\rm NN}}= 2.4$~GeV was of the order of 20\% \cite{Harabasz:2020sei}. However, rapidity distributions turned out to be systematically narrower in the model than in the experiment. This indicated that our assumption of the spherical symmetry was not exactly fulfilled, at least in the momentum space.

In this work, we extend our approach by allowing for spheroidal symmetry of the system, parameterized by the two eccentricities in the longitudinal (``beam'', $z$) direction, separately in the momentum and the position space. For a more systematic study, we present the results obtained with the two sets of chemical freeze-out parameters, as found in Refs.~\cite{Harabasz:2020sei} and~\cite{Motornenko:2021nds}. They are denoted below as Case A and Case B, respectively. As before, we select a reaction centrality class where thermalization is most likely to occur, i.e., we consider the 10\% most central collisions only. The decay of $\Delta(1232)$ is included via density function obtained in Ref.~\cite{Lo:2017ldt} from the pion-nucleon phase shift analysis. In Case B the same excited nuclear states are included in the present calculations as those used in Ref.~\cite{Motornenko:2021nds}.

\section{Spheroidal Siemens-Rasmussen model}  
As in our previous study \cite{Harabasz:2020sei}, the basis for this model remains the Cooper-Frye formula~\cite{Cooper:1974mv} that describes the invariant momentum spectrum of particles emitted from an expanding source,
\begin{equation}
E_p \frac{dN}{d^3p} = \int d^3\Sigma(x) \cdot p \,\, f(x,p).
\label{eq:CF1}
\end{equation}
Here $f$ is the phase-space distribution function of particles, $p$ is their four-momentum with the mass-shell energy,  $p^0 = E_p = \sqrt{m^2 + \boldsymbol{p}^2}$, and $d^3\Sigma_\mu(x)$ is the element of a three-dimensional freeze-out hypersurface from which particles are emitted.\footnote{Three-vectors are shown in bold font (unless stated otherwise) and a dot is used to denote the scalar product of both four- and three-vectors. The metric convention is ``mostly minuses'', $g_{\mu\nu} = \hbox{diag}(+1,-1,-1,-1)$.} 

In the calculations of the total particle yields, one can exchange the order of performing the integrals over the momentum space and the freeze-out hypersurface, 
\begin{equation}
N = \int d^3\Sigma_\mu(x) \int \frac{d^3p}{E_p} p^\mu f(x,p). \label{eq:N1} 
\end{equation}
Since the equilibrium distribution function  depends on the product $p \cdot u$ and thermodynamic parameters only (which are assumed to be constant on the freeze-out hypersurface), see Eq.~(6) in Ref.~\cite{Harabasz:2020sei}, we can further write 
\begin{eqnarray}
N\!&=&\!n(T,\Upsilon) \int d^3\Sigma(x) \cdot u(x) \equiv n(T,\Upsilon) {\cal V}, \label{eq:N2}  \\\nn
\end{eqnarray}
where the invariant volume ${\cal V}=\int d^3\Sigma(x) \cdot u(x)$ is defined by the integral in the middle part of Eq.~(\ref{eq:N2}). The fugacity $\Upsilon$ is defined as~\cite{Torrieri:2004zz}
\begin{eqnarray}
\Upsilon=  
 \gamma_q^{N_q+N_{\bar q}} \gamma_s^{N_s+N_{\bar s}}  \exp \left( \frac{\mu}{T}\right),
\label{upsiNeq}
\end{eqnarray}
We note that in the studies of the ratios of hadronic abundances, the invariant volume cancels out.

We modify the hitherto spherically-symmetric SR model by allowing the system to be stretched or squeezed in the beam direction. This is taken into account by the two eccentricity parameters: $\delta$ (for the momentum space) and $\epsilon$ (for the position space). Then, the space-time points lying on the freeze-out hypersurface have the following parametrization
\begin{equation}
x^\mu = \LR t, r \sqrt{1-\epsilon}\sin\theta\,\mathbf{\hat{e}_\rho}, r \sqrt{1+\epsilon}\cos\theta \RR,
\label{eq:xspheroid} 
\end{equation}
where $\mathbf{\hat{e}_\rho}=(\cos\phi, \sin\phi)$, while $\phi$ and $\theta$ are azimuthal and polar angles, respectively. In order to completely specify the freeze-out hypersurface, a curve in the $t-r$ plane has to be defined by the mapping   $\zeta\rightarrow\left(t(\zeta),r(\zeta)\right)$. This curve defines the (freeze-out) times $t$ when the hadrons in the (spheroidal) shells of radius $r$ stop to interact. The range of $\zeta$ may be always restricted to the interval: $0 \leq \zeta \leq~1$. The resulting infinitesimal element of the spheroidal hypersurface has the form
\begin{widetext}
\begin{eqnarray}
\! d^3\Sigma_\mu \!&=& (1-\epsilon)\!\LR 
r^\prime  \sqrt{1+\epsilon}, 
t^\prime \frac{\sqrt{1+\epsilon}}{\sqrt{1-\epsilon}}\sin\theta\,\mathbf{\hat{e}_\rho},
t^\prime \cos\theta\RR
r^2  d\Omega\,d\zeta,
\label{eq:d3sigmaspheroid}
\end{eqnarray}
where $d\Omega = \sin\theta\,d\theta\,d\phi$ is an infinitesimal element of the solid angle and the prime denotes the derivatives taken with respect to $\zeta$. 
If we assume the \emph{instantaneous freeze-out}, $t^\prime=0$, and use the parametrization $\zeta=r$, Eq.~(\ref{eq:d3sigmaspheroid}) reduces to
\begin{eqnarray}
\! d^3\Sigma_\mu \!&=& (1-\epsilon)\!\LR 
\sqrt{1+\epsilon}, 0, 0, 0\RR
 d\Omega\,r^2 dr. 
\label{eq:d3sigmaspheroid_instant}
\end{eqnarray}
Besides the spheroidally-symmetric hypersurface, we introduce a spheroidally-symmetric hydrodynamic flow,
\begin{eqnarray}
u^\mu &=& \gamma(\zeta,\theta)\LR 1,
v(\zeta) \sqrt{1-\delta}\sin\theta\,\mathbf{\hat{e}_\rho}, 
v(\zeta) \sqrt{1+\delta}\cos\theta \RR, 
\label{eq:uspheroid}
\end{eqnarray}
\begin{figure*}
    \centering
    \includegraphics[width=\textwidth]{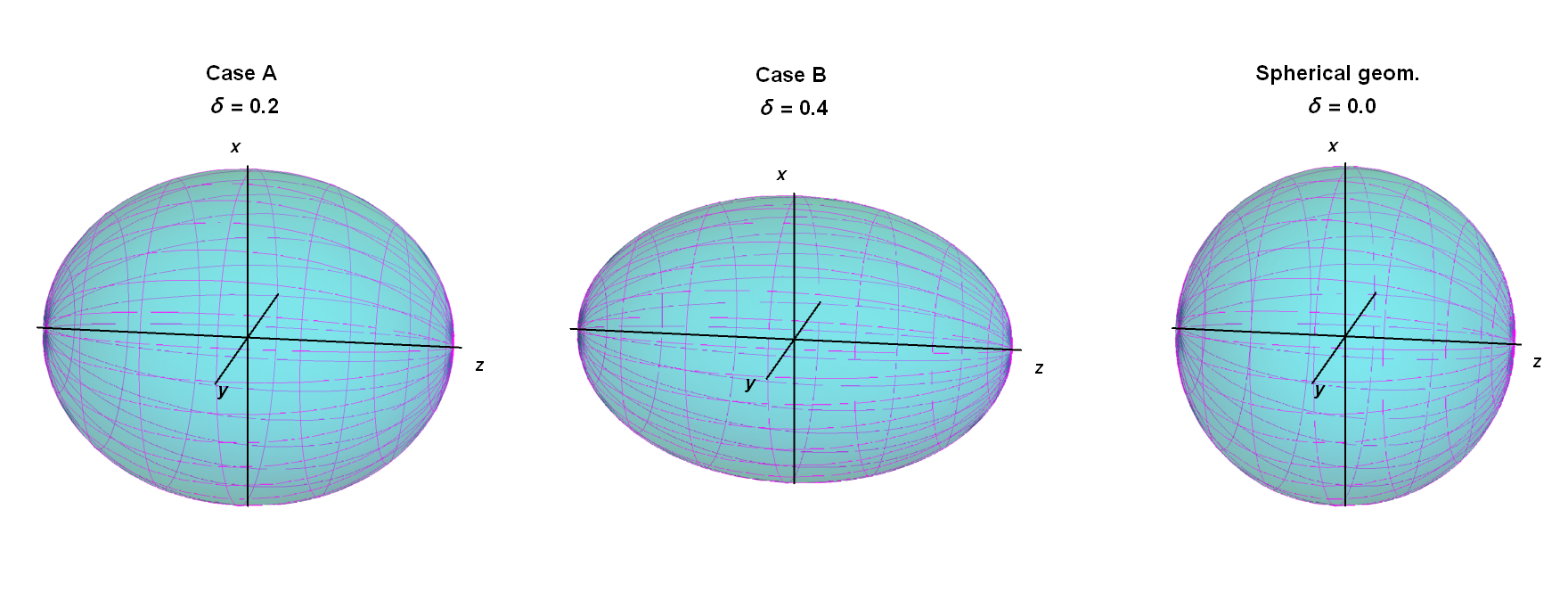}
    \caption{Graphical representation of the flow parametrization for the three studied cases. The points on the surfaces represent solutions of the equation $ (v_x^2 + v_y^2)/(1-\delta)+v_z^2/(1+\delta) = v^2$. }
    \label{fig:spheroids}
\end{figure*}
where $\gamma(\zeta,\theta)$ is the Lorentz factor that, due to the normalization condition $u\cdot u=1$, is given by the formula $\gamma(\zeta,\theta) = \LS 1-(1+\delta \cos(2 \theta))v^2(\zeta)\RS^{-1/2}$; for graphical representation of Eq.~(\ref{eq:uspheroid}) for the values of $\delta$ obtained in this analysis see Fig.~\ref{fig:spheroids}. Thus, we can write explicit expressions for the two dot products used in the numerical Monte-Carlo calculations, namely, the product of the four-velocity and four-momentum vectors
\begin{equation}
 u \cdot p=\gamma(\zeta,\theta) \LS E_p\!-\!p\, v(\zeta) \LR \sqrt{1+\delta}\cos\theta \cos\theta_p + \sqrt{1-\delta}\sin\theta \sin\theta_p \cos(\phi-\phi_p)\RR \RS,
\label{eq:puspheroid}
\end{equation}
where the subscript `$p$' denotes quantities related to the momentum vector, and the product of the hypersurface element and the four-momentum vector
\begin{equation}
d^3\Sigma \cdot p = (1-\epsilon)\LS E_p r^\prime \sqrt{1+\epsilon} 
- p\, t^\prime  \LR \cos\theta \cos\theta_p + \frac{\sqrt{1+\epsilon}}{\sqrt {1-\epsilon}} \sin\theta \sin\theta_p \cos(\phi-\phi_p)\RR \RS r^2 d\Omega \, d\zeta.
\label{eq:Sigmapspheroid}
\end{equation}
In the case of the instantaneous freeze-out introduced above, this expression simplifies to
\begin{equation}
d^3\Sigma \cdot p = (1-\epsilon) E_p \sqrt{1+\epsilon}\, d\Omega \, r^2 dr.
\label{eq:Sigmapspheroid_instant}
\end{equation}
The invariant volume defined by Eq.~(\ref{eq:N2}) takes the following form for the spheroidally-symmetric system
\begin{equation}
{\cal V} = (1-\epsilon)\int\, \gamma(\zeta,\theta) \LC r^\prime 
\sqrt{1+\epsilon}
- t^\prime v(\zeta)  \frac{\sqrt{1-\delta}}{2}  \LS 
\frac{\sqrt{1+\delta}}{\sqrt{1-\delta}} 
+\frac{\sqrt{1+\epsilon}}{\sqrt{1-\epsilon}} +\LR  
\frac{\sqrt{1+\delta}}{\sqrt {1-\delta}} 
-\frac{\sqrt{1+\epsilon}}{\sqrt {1-\epsilon}}\RR  \cos \LR 2\theta 
\RR  \RS \RC \, r^2 d\Omega \, d\zeta,
\label{eq:Sigmauspheroid}
\end{equation}
\end{widetext}
which in the case of the instantaneous freeze-out and the Hubble-like radial profile of the flow velocity ($v=\tanh(Hr)$, where $H$ is a constant parameter \cite{Chojnacki:2004ec}) reduces to
\begin{equation}
{\cal V} = (1-\epsilon)\int_0^R\int\, \gamma(\zeta,\theta)  
\sqrt{1+\epsilon}\, d\Omega \, r^2 dr.
\label{eq:Sigmauspheroid_instant}
\end{equation}
Here the parameter $R$ is fixed by the abundance of one of the particle species (note that the abundances of the other species are known through the ratios set by thermodynamic parameters). For $\epsilon~=~\delta~=~0$, all the above equations reduce to the standard, spherically-symmetric version of the SR model used in Ref.~\cite{Harabasz:2020sei}.
%

\section{Thermal parameters}

Input parameters for our calculations were obtained by different fitting strategies of the particle multiplicities calculated in the grand-canonical ensemble to the ones measured by the HADES collaboration in Au-Au collisions at $\sqrt{s_{\rm NN}}= 2.4$~GeV, as listed in Table \ref{tab:mult}.
As a result of the fit, we consider two sets of thermodynamic parameters whose values are listed in the upper section of Table \ref{tab:params}. 
\begin{table}[b]
\begin{center}
\begingroup
\setlength{\tabcolsep}{7.5pt} 
\renewcommand{\arraystretch}{1.2} 
\begin{tabular}{ |CCCC| }
\hline \hline
    \text{particle}  & \text{multiplicity} \:\:\: & \text{uncertainty} & \text{Ref.} \\ \hline 
    p  \: \text{(free)}     & 77.6   & \pm 2.4 & \text{\cite{Szala:2019,Szala2019a}}\\
    p+n\rightarrow {}^2\mathrm{H} & 28.7 &  \pm  0.8 &\text{\cite{Szala:2019,Szala2019a}}\\
    p+2n\rightarrow {}^3\mathrm{H} & 8.7 &  \pm  1.1 &\text{\cite{Szala:2019,Szala2019a}}\\
    2p+n\rightarrow {}^3\mathrm{He} & 4.6 &  \pm  0.3 &\text{\cite{Szala:2019,Szala2019a}}\\
    p \: \text{(bound)} & 46.5 &  \pm  1.5 &\text{\cite{Szala:2019,Szala2019a}}\\
    \pi^+  & 9.3   &  \pm 0.6 & \text{\cite{HADES:2020ver}}\\
    \pi^-  & 17.1   & \pm 1.1 & \text{\cite{HADES:2020ver}}\\
    K^+    & 5.98\,10^{-2} & \pm 6.79\,10^{-3}  & \text{\cite{Adamczewski-Musch:2017rtf}} \\
    K^-    & 5.6\,10^{-4} & \pm 5.96 \,10^{-5} & \text{\cite{Adamczewski-Musch:2017rtf}} \\
    \Lambda & 8.22\,10^{-2} & ^{+5.2} _{-9.2}\,10^{-3} & \text{\cite{Adamczewski-Musch:2018xwg}} \\ 
    \hline \hline
\end{tabular}
\endgroup
\caption{Particle multiplicities used in the determination of the freeze-out parameters. Protons bound in nuclei are taken into account as shown. \label{tab:mult}}
\end{center}
\end{table}

In the first case (dubbed below as Case A), the calculation method and resulting thermodynamic parameters are the same as in Ref.~\cite{Harabasz:2020sei}. The number of free parameters equals the number of available particle yields. As the total number of protons we take the sum of those observed directly in the experiment and those observed as bound in detected light nuclei.

The second case (Case B) has been elaborated in Ref.~\cite{Motornenko:2021nds}. The same experimental data and the same approach to proton counting have been used as in Ref.~\cite{Harabasz:2020sei}. However, additional constraints of $S=0$ and $Q/B=0.4$ have been applied to the total values of the conserved quantum numbers, which reduced the number of free parameters by two. This allows to calculate the $\chi^2$ of the fit, in contrast to Case A where the number of degrees of freedom is 0. The $\chi^2$ profile turns out to have two nearly degenerate minima. The first one corresponds to thermodynamic parameters nearly identical to those found in Case A, so we do not consider it separately. The second minimum has resulted in substantially higher temperature and chemical potential but smaller fireball size, this defines Case B.


\begin{table*}[t]
    \centering
    \begingroup
    \setlength{\tabcolsep}{13pt} 
    \renewcommand{\arraystretch}{1.2} 
    \begin{tabular}{|c|c|c|c|}
    \hline
    \hline
    Parameter & \makecell{Spherical geometry, \\Ref. \cite{Harabasz:2020sei}} & Case A & Case B \\
    \hline
    $T$ (MeV) & $49.6$ & $49.6$ & $70.3$ \\
    $R$ (fm) & $ 16.0 $  & $ 15.7 $ & $6.06$ \\
    $\mu_B$ (MeV) & $776$  & $776$ & $876$ \\
    $\mu_S$ (MeV) & $123.4$  & $123.4$ & $198.3$ \\
    $\mu_{I_3}$ (MeV) & $-14.1$  & $-14.1$ & $-21.5$ \\
    $\gamma_S$ & $0.16$ &  $0.16$ & $0.05$ \\
    $\chi^2/N_{\rm df}$ & $ N_{\rm df}=0 $  & $ N_{\rm df}=0 $ & $1.13/2$ \\
    \hline
    \hline
    $H$ (GeV) & $0.008$  & $0.01$ & $0.0225$ \\
    $\delta$ & $0$  & $0.2$ & $0.4$ \\
    $\sqrt{Q^2}$ & $0.285$ & $0.238$ & $0.256$ \\
    \hline
    \hline
    \end{tabular}
    \endgroup
    \caption{Upper part: thermodynamic parameters obtained from different fitting strategies to ratios of particle multiplicities measured in Au-Au collisions at $\sqrt{s_{_{NN}}}=2.4$ GeV. Lower part: Hubble-like expansion parameter $H$ and momentum-space longitudinal eccentricity $\delta$ obtained from fitting to the experimentally measured protons and pions spectra.}
    \label{tab:params}
\end{table*}

\section{Particle interferometry}
By measuring quantum-mechanical correlations between particles emitted from a given source, it is possible to estimate the spatial extension of the source (or at least of homogeneity regions within it).
It is beneficial for the sources which cannot be measured by other means, like distant stars \cite{HanburyBrown:1956bqd} or hadron sources in heavy-ion collisions \cite{Shuryak:1972kq}.

We study the two-pion interferometry in analogy to Ref.~\cite{Kisiel:2006is}.
The main object of interest is the two-particle correlation function, which is defined in general as
\begin{equation}
    {\cal C}(p_1,p_2) = \frac{W_2(p_1,p_2)}{W_1(p_1)W_1(p_2)},
\end{equation}
where one- and two-particle emission functions are defined as follows
\begin{eqnarray}
W_1(p_1) &=& E_{p_1}\frac{dN}{d^3p_1} \nn \\
&=&\int S(x_1,p_1)\,d^4x_1, \nn \\
W_2(p_1,p_2) &=& E_{p_1}E_{p_2}\frac{dN}{d^3p_1 d^3p_2} \nn \\
&=&\int S(x_1,x_2,p_1,p_2)\,d^4x_1d^4x_2. \nn
\end{eqnarray}
Here $S(x_1,p_1)$ and $S(x_1,x_2,p_1,p_2)$ are one- and two-particle distributions in position and momentum spaces. Making the \textit{smoothness approximation}, the correlation function can be written as
\begin{equation}
    C(q,k)=\frac{\int S(x_1,p_1)S(x_2,p_2)|\Psi(k^*,r^*)|^2d^4x_1d^4x_2}{\int S(x_1,p_1)\,d^4x_1\int S(x_2,p_2)\,d^4x_2},
    \label{eq:cf_smoothness}
\end{equation}
where the wave function of two particles $\Psi(k^*,r^*)$ has been introduced, depending on their relative momentum $k^*$ and position $r^*$ in the pair rest frame. Furthermore, $q = (E_{p_1}-E_{p_2}, p_1 - p_2)$ denotes the momentum difference between the two particles and $k=\frac{1}{2}(E_{p_1} + E_{p_2}, p_1 + p_2)$ is the average momentum of the pair.

THERMINATOR generates primordial particles independently. Even though the Bose-Einstein or Fermi-Dirac statistics enters the particle distribution function $f(x,p)$ in Eq.~(\ref{eq:CF1}), particles are classical, having well defined positions and momenta. In order to introduce quantum-mechanical correlations, we consider the wave function of two non-interacting particles
\begin{equation}
\Psi(k^*,r^*)=\frac{1}{\sqrt{2}}(e^{ik^*r^*} + e^{-ik^*r^*}).
\label{eq:piWaveFunc}
\end{equation}
Its square, which enters the correlation function, reads
\[
|\Psi(k^*,r^*)|^2 = 1+\cos(2k^*r^*).
\]
In the Monte-Carlo simulation, particles are grouped into events, as in the experiment, and the integrals in Eq.~(\ref{eq:cf_smoothness}) are represented by sums over all 2-particle combinations in a given event. This is realised by assigning pairs to bins, which can be formally expressed with the help of the function
\begin{equation}
    \delta_\Delta(x)=
    \begin{cases}
    1 & \text{if } |x|\leq\frac{\Delta}{2}, \\
    0 & \text{otherwise}.
    \end{cases}
\end{equation}
The correlation function then takes the form
\begin{widetext}
\begin{equation}
    C(q,k)=\frac{\sum_i\sum_{j\neq i} \delta_\Delta(q-p_i+p_j)\delta_\Delta(k-\frac{1}{2}(p_i+p_j))|\Psi(k^*,r^*)|^2}{\sum_{i}\sum_{j}\, \delta_\Delta(q-p_i+p_j)\delta_\Delta(k-\frac{1}{2}(p_i+p_j))}.
    \label{eq:corrTherminator}
\end{equation}
\end{widetext}
This allows to build histograms of the usual momentum projections $q_\text{inv}$ for the one-dimensional representation or $q_\text{out}$, $q_\text{side}$, $q_\text{long}$ for the three-dimensional Bertsch-Pratt decomposition \cite{Pratt:1986cc}.  The size of the homogeneity region in the fireball is then estimated by reciprocals of widths of the respective distributions. Assuming pion wave function as defined in Eq.~(\ref{eq:piWaveFunc}), and the single-particle emission function to be a three-dimensional ellipsoid with three parallel components: \textit{out}, \textit{side} and \textit{long}, Eq.~(\ref{eq:cf_smoothness}) leads to the formula
\begin{eqnarray}
    C(q,k) = 1 + \lambda_\text{osl} e^{-R^2_\text{out}q^2_\text{out} -R^2_\text{side}q^2_\text{side} -R^2_\text{long}q^2_\text{long}}.
    \label{eq:corrFit3D}
\end{eqnarray}

In the case of a static, spherically-symmetric system one may obtain
\begin{equation}
     C(q,k) = 1 + \lambda_\text{inv} e^{-R^2_\text{inv}q^2_\text{inv}}.
     \label{eq:corrFit1D}
\end{equation}
The parameters $R_\text{out}$, $R_\text{side}$, $R_\text{long}$ and $R_\text{inv}$ represent the mentioned widths of Gaussian approximation of the fireball and are often referred to as the \textit{HBT radii}. 
Equations (\ref{eq:corrFit3D}) and (\ref{eq:corrFit1D}) are commonly used to fit experimental data or, in our case, the generated data of a Monte-Carlo event generator, given by Eq.~(\ref{eq:corrTherminator}). 
The correlation strength of pairs of particles can be tuned with two independent parameters $\lambda_\text{inv}$ and $\lambda_\text{osl}$.
In our analysis, they are stay constant as functions of the average transverse pair momentum. 


\section{Fitting procedure}

The parameters $H$, $\epsilon$, and $\delta$ are adjusted to achieve the best agreement between the model and the experimental data. The agreement is quantified by calculating the mean relative error
\begin{equation}
Q = \sqrt{\frac{1}{N}\sum_{i=1}^{N}\frac{\left(Y_{i,\rm{model}}-Y_{i,\rm{exp}}\right)^2}{Y^2_{i,\rm{exp}}}},
\label{eq:mean_rel_error}
\end{equation}
where the sum runs over all the bins of histograms of experimental and model data, which were taken for comparison, while $Y_{i,\rm{exp}}$ and $Y_{i,\rm{model}}$ are experimental and model results in these bins, respectively. In this work, we take for comparison transverse mass distributions of protons, $\pi^+$'s and $\pi^-$'s in five center-of-mass rapidity intervals: $[-0.45, -0.35]$, $[-0.25, -0.15]$, $[-0.05, 0.05]$, $[0.15, 0,25]$ and $[0.35, 0.45]$, giving in total $N=253$ bins.

We have found that values of $Q$ (and the model transverse mass distributions themselves) are practically independent of $\epsilon$. Therefore, we keep $\epsilon=0$ and perform a simple grid search in a $H-\delta$ plane to find the values of these parameters which minimize $Q$. For each point of the grid, we adjust $R$ to keep ${\cal V}$ in Eq.~(\ref{eq:Sigmauspheroid_instant}) unchanged, which makes the particle yields the same. In the next step we keep $H$ and $\delta$ fixed and compare the HBT radii between the model and the experiment for different values of $\epsilon$. 

The results of the grid scan are shown in the lower section of Table~\ref{tab:params} for the considered two sets of thermodynamic parameters. They are also compared to the spherically-symmetric case with $H = 0.008$~GeV and $\delta=0$ taken from Ref. \cite{Harabasz:2020sei}. The mean relative error is lowest in Case A, $Q = 0.238$, slightly larger for Case B, $Q = 0.256$, and the highest for the spherical version, $Q = 0.285$.

One may notice that the values of $Q$ obtained in the cases A and B are very close to $Q$ reported previously in Ref.~\cite{Harabasz:2020sei} and argue that the generalization of the model from spherical to spheroidal symmetry would not bring significant improvement. However, in our previous work a different set of data points was used to calculate $Q$: rapidity distributions and transverse mass spectra at mid-rapidity for $p$, $\pi^+$ and $\pi^-$. In the present work, we use the transverse mass spectra in five rapidity ranges. In this way, for the spherical case we obtain $Q=0.285$, a significantly higher value than those found for the spheroidal fireball.
\begin{figure*}
    \centering
    \includegraphics[width=\textwidth]{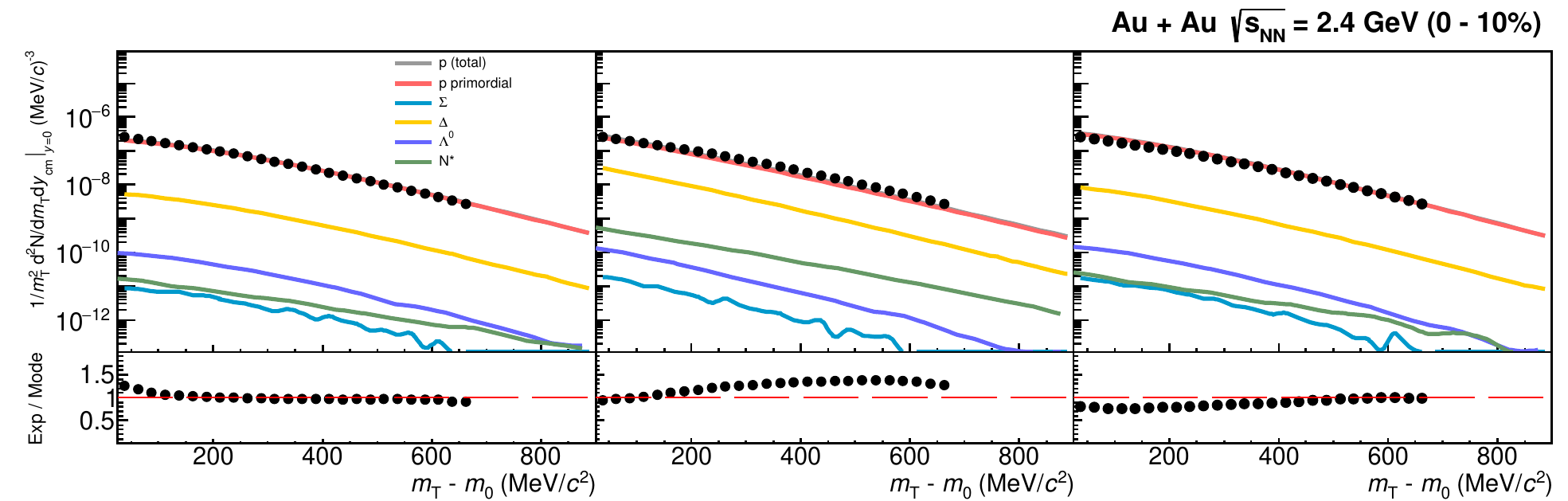}
    \caption{Mid-rapidity transverse-mass distributions of protons measured in Au-Au collisions at $\sqrt{s_{\rm NN}} = 2.4$~GeV, compared to the model calculations done for the three different strategies of obtaining thermodynamic parameters: Case A (left), Case B (middle), and spherical geometry (right). Colored lines show different contributions as explained in the legend. Ratios of experimental data to the model are shown in the lower panels.}
    \label{fig:proton_mt}
\end{figure*}
\begin{figure*}
    \centering
    \includegraphics[width=\textwidth]{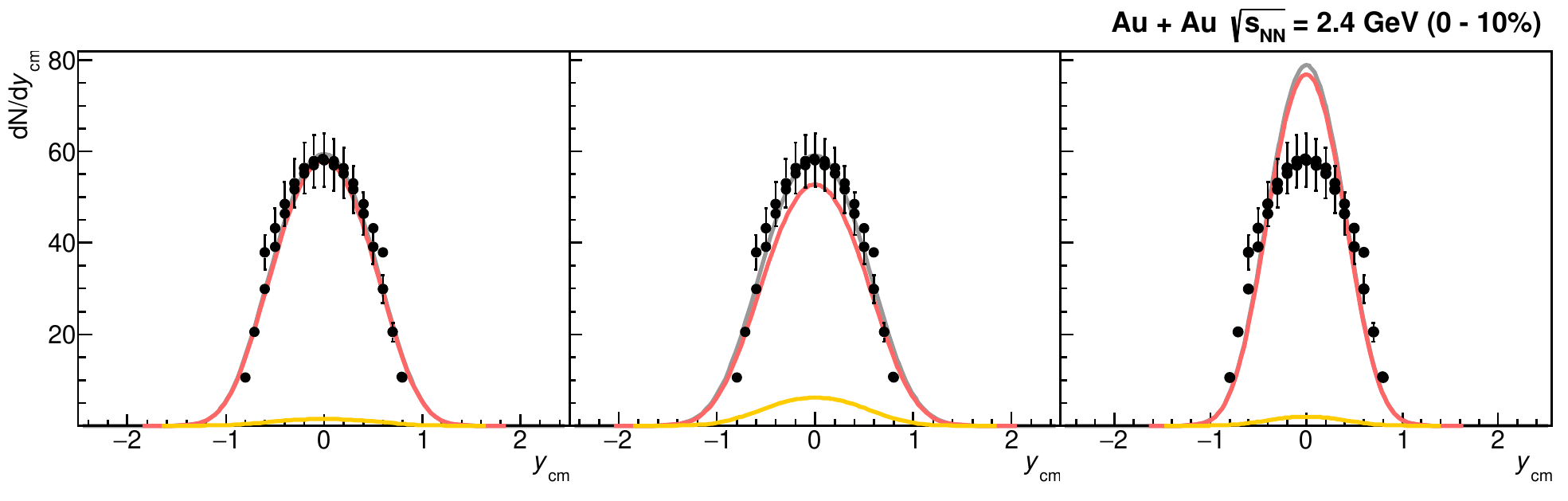}
    \caption{Rapidity distributions of protons measured in Au-Au collisions at $\sqrt{s_{\rm NN}} = 2.4$~GeV, compared to the model calculations done for the three different strategies of obtaining thermodynamic parameters: Case A (left), B (middle), and spherical geometry (right). Colored lines show different contributions as explained in the legend.}
    \label{fig:proton_y}
\end{figure*}
\begin{figure*}
    \centering
    \includegraphics[width=\textwidth]{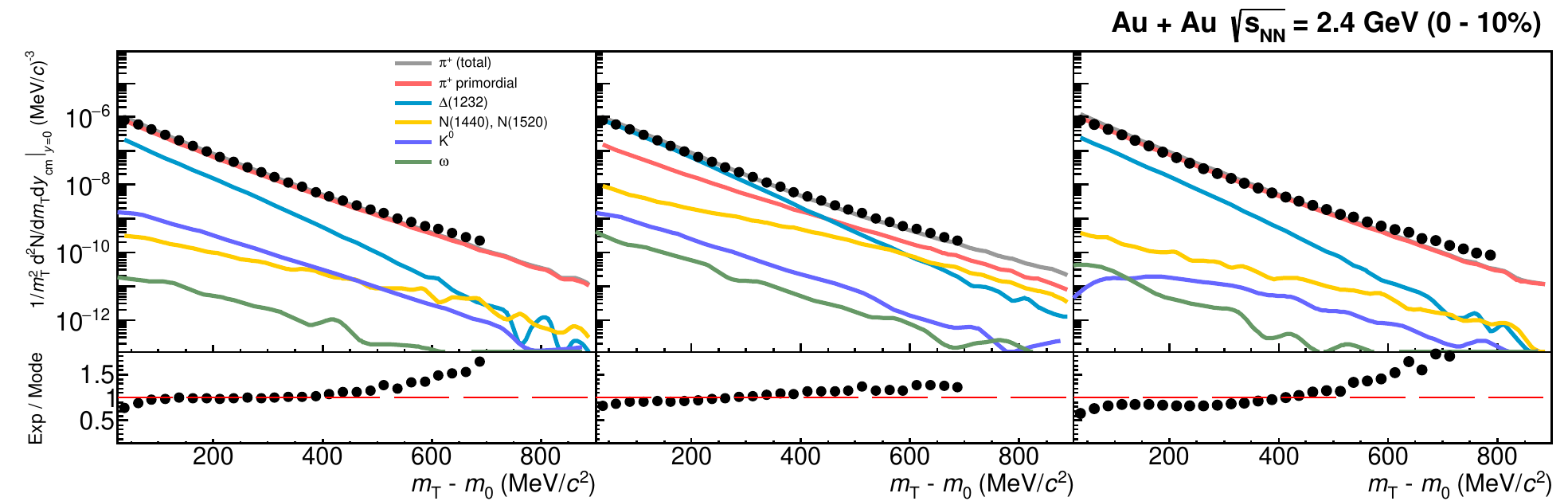}
    \caption{Same as Fig. \ref{fig:proton_mt} but for positively charged pions.}
    \label{fig:pip_mt}
\end{figure*}
\begin{figure*}
    \centering
    \includegraphics[width=\textwidth]{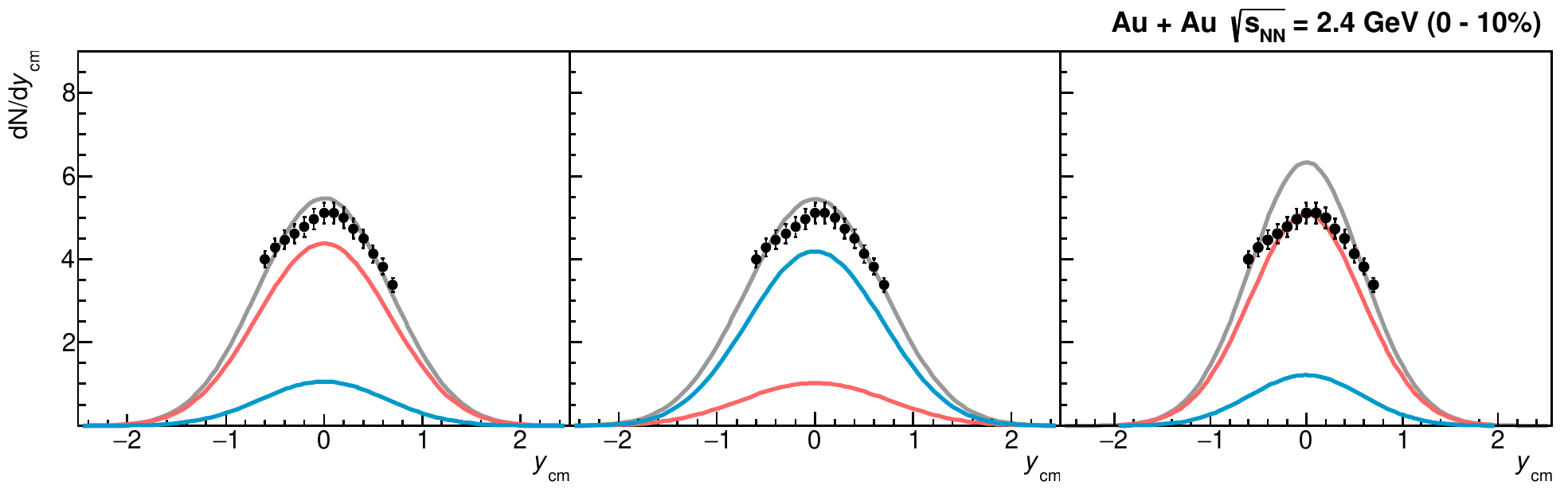}
    \caption{Same as Fig. \ref{fig:proton_y} but for positively charged pions.}
    \label{fig:pip_y}
\end{figure*}
\begin{figure*}
    \centering
    \includegraphics[width=\textwidth]{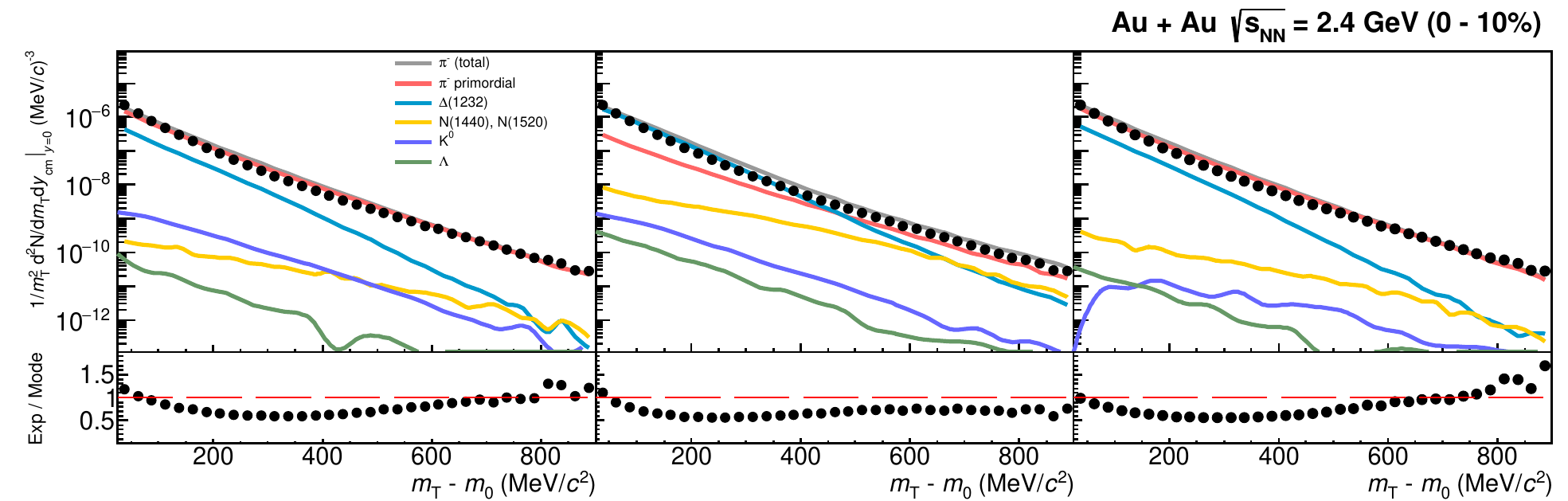}
    \caption{Same as Fig. \ref{fig:proton_mt} but for negatively charged pions.}
    \label{fig:pim_mt}
\end{figure*}
\begin{figure*}
    \centering
    \includegraphics[width=\textwidth]{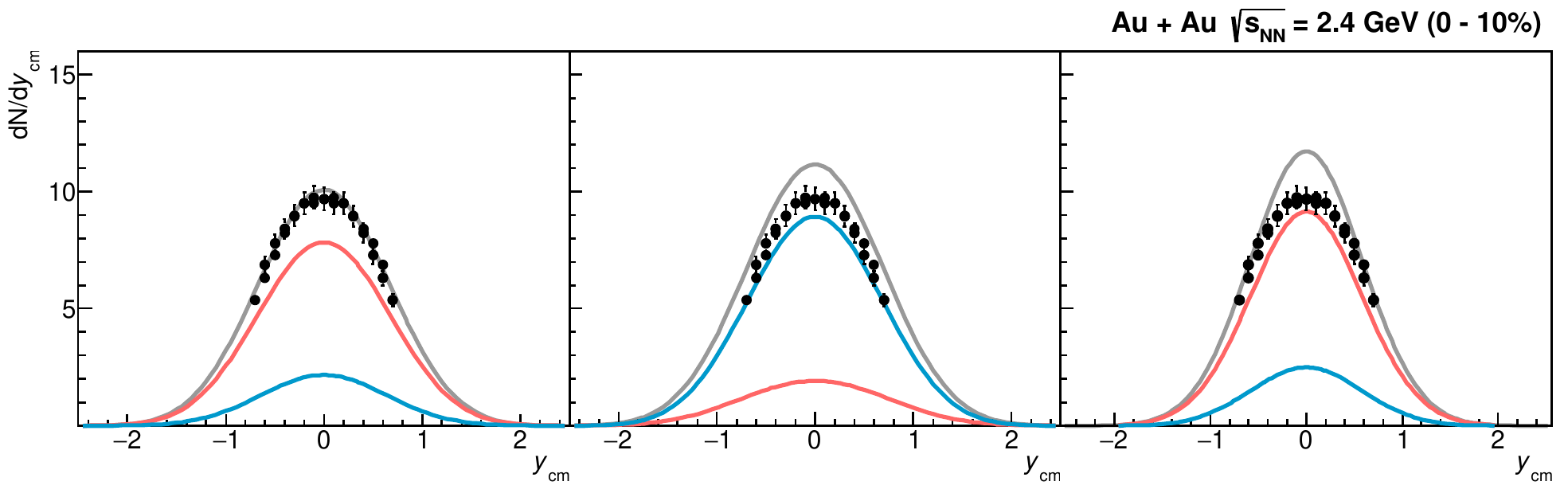}
    \caption{Same as Fig. \ref{fig:proton_y} but for negatively charged pions.}
    \label{fig:pim_y}
\end{figure*}

\section{Results and discussion}
\subsection{Inclusive spectra of protons and pions}

Figures \ref{fig:proton_mt}--\ref{fig:pim_y} show the comparison of the model transverse-mass (at mid-rapidity) and rapidity distributions with the ones measured by HADES in Au-Au collisions at $\sqrt{s_{\rm NN}}= 2.4$~GeV. The left, middle, and right panels show the results for Case A, Case B, and the spherical model, respectively. The obtained results are presented for protons, Figs.~\ref{fig:proton_mt}-\ref{fig:proton_y}, positive pions, Figs.~\ref{fig:pip_mt}-\ref{fig:pip_y}, and negative pions Figs.~\ref{fig:pim_mt}-\ref{fig:pim_y}. 

Lines with different colors show different contributions to the total spectra (shown in gray). The most important contribution (shown in red) is from ``primordial'' particles -- they are directly generated on the freeze-out hypersurface. Other contributions are from feed-down, i.e., from unstable particles generated on the freeze-out hypersurface and subsequently decaying into protons or pions. At temperatures reach in HIC energies considered here, the only decay which significantly contributes  to total particle multiplicities is that from the $\Delta(1232)$ resonances. In Case B, also $N^*(1440)$ and $N^*(1520)$ play a role at large transverse-masses of pions.

If the freeze-out temperature is relatively low, as in Case A, the dominating contribution to pions comes from primordial particles. Their distribution in the representation $dN/(m_T^2 \, dm_T)$ is slightly convex in the region where $m_T-m_0>200~\text{MeV}$ due to the collective radial expansion. Without such expansion, they would have the form of the Bose-Einstein distributions, which in this transverse-mass region are practically equal to the Boltzmann distributions represented by the exponential functions. Anyway, the curvature of the model spectra is slightly smaller than that observed in the data.

The situation is different in Case B. 
At low transverse mass, the yield is dominated by the $\Delta(1232)$ contribution which is relatively steep. At large transverse mass, the main contributions come from primordial pions and $N^*$ resonances. Their slope is smaller, in particular, for $N^*$. Combined contributions quite well reproduce the spectra, especially for $\pi^+$. On the other hand, the experimental spectrum of $\pi^-$ bends up much stronger than the model curve, while moving from high to low values of $m_T-m_0$. 

This effect could be attributed to the electromagnetic interaction with the positively charged fireball. In fact, in all three cases, one can see such a trend at low transverse mass. The ratios of experimental data to the model bend up for $\pi^-$ and bend down, somewhat less strongly, for $\pi^+$. Final-state electromagnetic interactions have not yet been implemented in our Monte-Carlo simulations.

Finally, we emphasize that both models A and B describe data, particularly rapidity distributions,  better than the model with spherical geometry.  
\subsection{Two-pion interferometry}

In the last sections we have showed that for the two considered sets of thermodynamic parameters, the optimal values of $H$ and $\delta$ give a similar level of agreement between the HADES experimental data and the model. Moreover, the inclusive hadron spectra discussed before are not sensitive to the position-space eccentricity of the fireball shape in the longitudinal direction. 

On the other hand, the size of the fireball is negatively correlated with temperature. A higher temperature leads to a higher particle density due to the Boltzmann factor, and one has to compensate for that with a smaller fireball size to obtain the particle multiplicities as observed in the experiment. This can be observed in Table \ref{tab:params}. Consequently, the measurements of the HBT radii, which are correlated with the size of the fireball, may allow for discriminating between the two sets of thermodynamic parameters. In addition, it can be expected that the HBT radii and the relations between them can depend on the parameter $\epsilon$ which quantifies the spatial eccentricity. 

\begin{figure*}
    \centering
    \includegraphics[width=.6\textwidth]{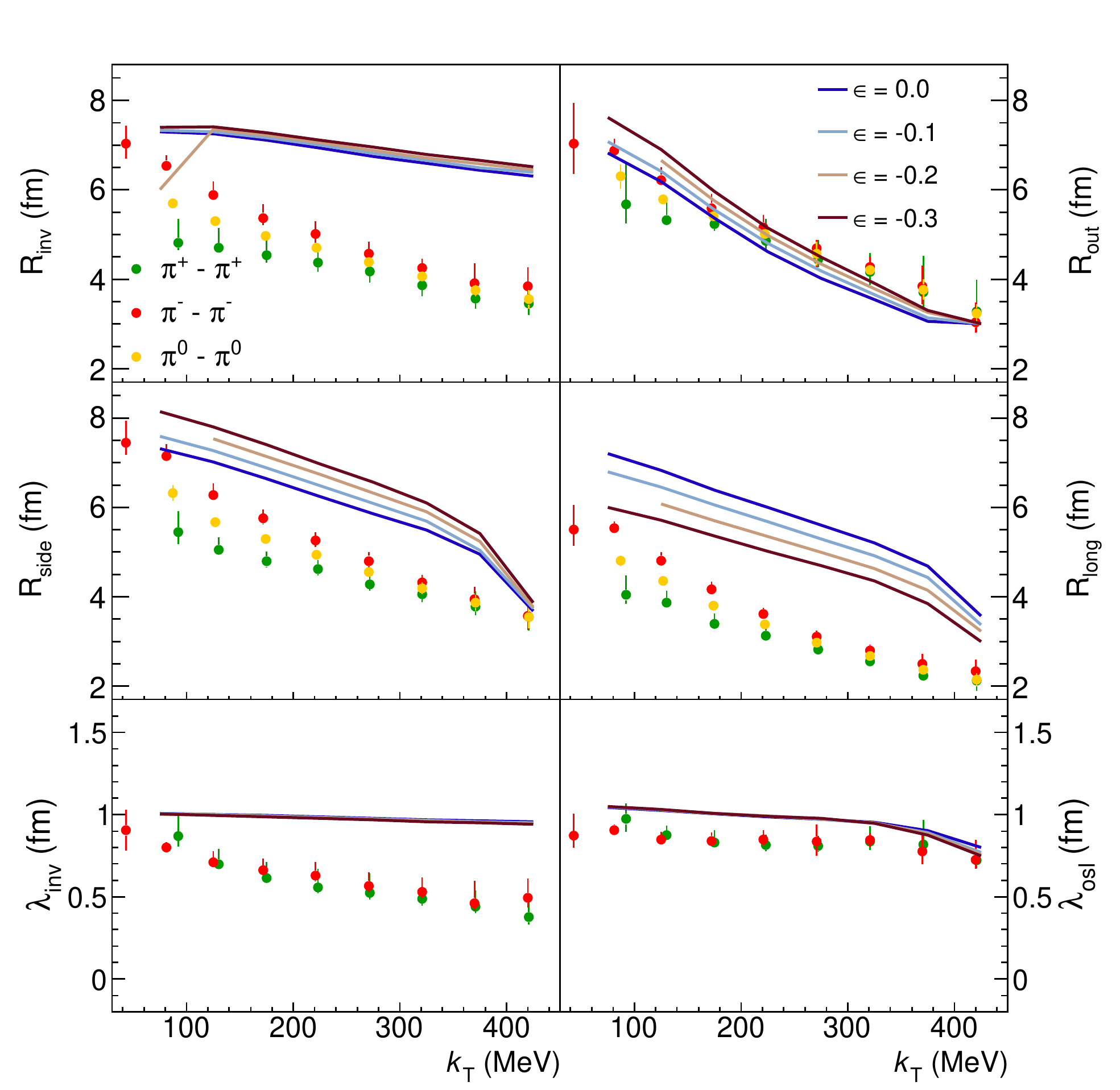}
    \caption{Emission source radii (lines) in the \textit{out}, \textit{side} and \textit{long} directions, as alongside invariant radius, for the Case A. Experimental results (points) of the HBT radii obtained form $\pi^+$, $\pi^-$ and $\pi^0$ correlation functions.}
    \label{fig:3dcorrA}
\end{figure*}

\begin{figure*}
    \centering
    \includegraphics[width=.6\textwidth]{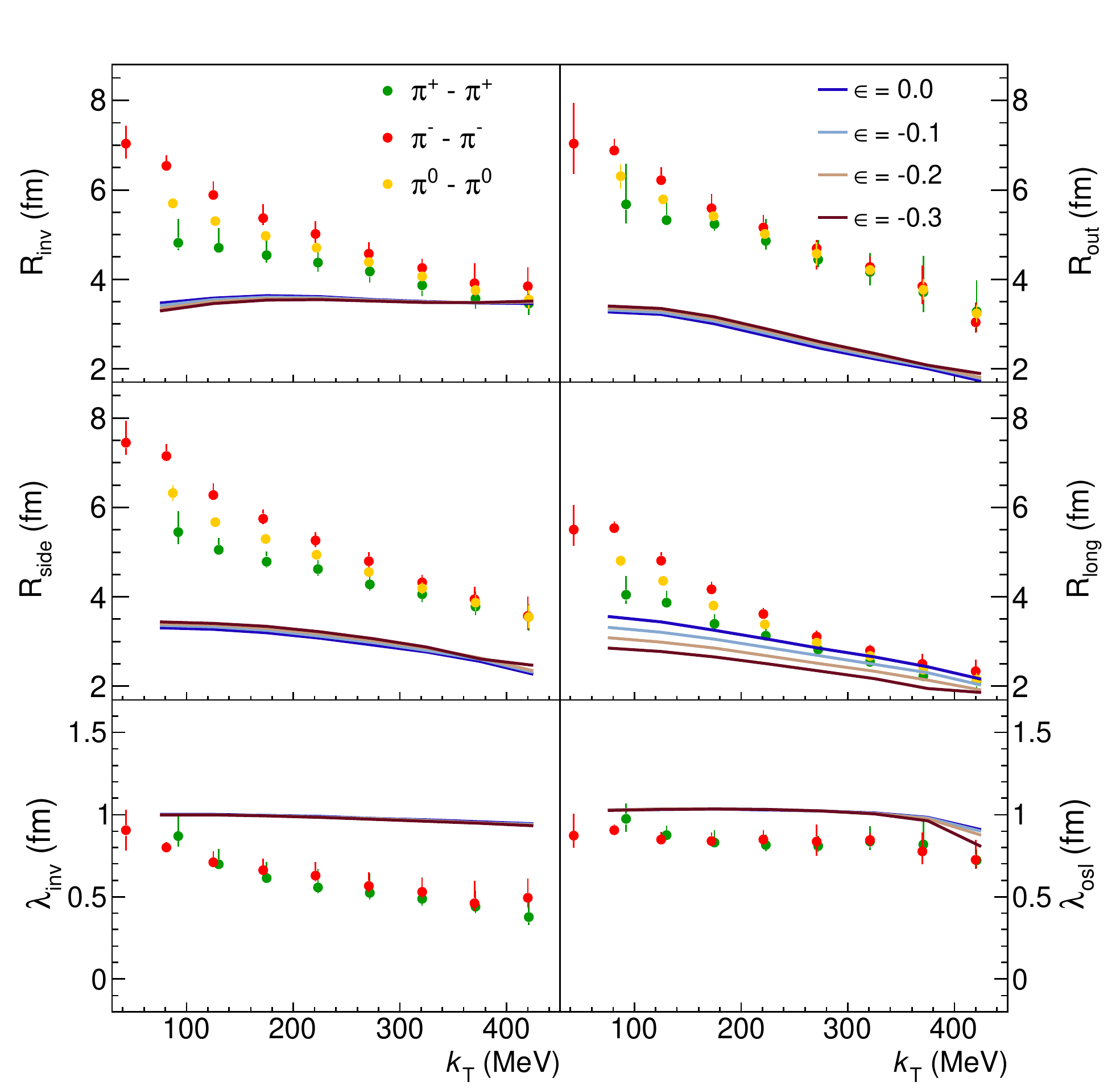}
    \caption{Same as Fig. \ref{fig:3dcorrA} but for the Case B. 
    }
    \label{fig:3dcorrB}
\end{figure*}

The above remarks motivated our additional study of the two-pion interferometry within the spheroidal model. The results of our calculations are shown in Figs. \ref{fig:3dcorrA} and \ref{fig:3dcorrB}, where we compare the model results with the HADES HBT data~\cite{HADES:2018gop}.

In contrast to the spectra, in the case of the correlation functions we obtain only qualitative agreement with the data. In Case A, the three model radii, $R_{\rm side}$, $R_{\rm out}$ and $R_{\rm long}$, strongly decrease with the mean transverse momentum of the pion pair, $k_T$, an effect that can be attributed to the presence of the large transverse flow. The model slopes are similar to those observed in the data, however, the values of the three model radii are larger or similar to the experimental values. In Case B the situation is opposite, the HBT radii and slopes are smaller. This behavior reflects the smaller size of the system (due to a higher temperature) and smaller transverse flow.

As expected, our results indicate dependence of the results on the anisotropy parameter $\epsilon$. However, none of the considered values of $\epsilon$ gives a good quantitative description of the data. The comparison of $R_{\rm long}$ with the data suggests large negative values of $\epsilon$, while the comparison of $R_{\rm side}$ and $R_{\rm out}$ suggests the values close to zero.

We note that the fit results presented in Figs.~\ref{fig:3dcorrA} and \ref{fig:3dcorrB} include cut on the $\lambda$ parameter of the correlation function to exclude cases where no positive fit result was obtained, $\lambda_{\rm inv/osl} \in (0.5,2)$.


We close our HBT analysis with the conclusion that the HBT modeling for our model requires further improvements. Theoretical analyses of the HBT radii in heavy-ion collisions studied at RHIC energies have showed that a successful description of the data can be achieved only if many improvements in the theoretical description of hadron emission are made~\cite{Pratt:2008qv}. They include, in particular, the use of realistic wave functions, viscous corrections to the distribution functions and inclusion of the hadron emission times. One way of improvement of our results is to make similar advances in our framework. The other way for achieving a better theoretical description of the data is to make global fits that include both one- and two-particle observables, i.e., to include the HBT results in the fitting procedure. 

\section{Summary and Conclusions} 

In this work, we have studied the rapidity and transverse-mass spectra of protons and pions produced in Au-Au collisions at $\sqrt{s_{\rm NN}} = 2.4$~GeV within a thermal model with single freeze-out. We have generalized the original, spherically-symmetric Siemens-Rasmussen approach by allowing for elongation or contraction of the fireball in the longitudinal direction, separately in the position and the momentum space. The momentum-space elongation allowed for reproducing the width of the experimentally measured rapidity distributions and significantly improved the agreement between the model and experimental data, as compared to spherical-model predictions. We have also found that the inclusive hadron spectra are not sensitive to the position-space shape of the fireball. With the model parameters fixed by the spectra we have calculated the HBT radii, which turned out to be only in qualitative agreement with the data. 

Altogether, our results bring evidence for substantial thermalization of the matter produced in the few-GeV energy range and its spheroidal expansion. This result is quite appealing and seems to be natural as in the considered energy range we expect stopping of colliding nucleons. 
This stopping, however, cannot be complete and we expect a residual imbalance in favour of longitudinal momentum.
This picture is supported by our calculations. 

Our findings pave the way for future developments of our approach that may include more realistic description of correlations and determination of the yields of light nuclei. A very natural extension would be to include also the asymmetry of particle emission in the transverse plane, i.e., to generalize the spheroidal symmetry to the ellipsoidal one. This could allow for studies of the elliptic flow of particles.

\section{Acknowledgements} This work was supported in part by:
the Polish National Science Center Grants  No.~2018/30/E/ST2/00432, No.~2017/26/M/ST2/00600, No.~2020/38/E/ST2/00019 and No.~2021/41/B/ST2/02409; IDUB-POB-FWEiTE-3,  project granted by Warsaw University of Technology under the program Excellence Initiative: Research University (ID-UB); TU Darmstadt, Darmstadt (Germany): HFHF, ELEMENTS:500/10.006, GSI F\&E, DAAD PPP Polen 2018/57393092;
Goethe-University, Frankfurt(Germany): HFHF, ELEMENTS:500/10.006.
JK acknowledges the scholarship grant from the GET$\_$INvolved Programme of FAIR/GSI.

\bibliography{main}{}

\end{document}